\def\<<{{\ll}}
\def\>>{{\gg}}
\def\spose#1{\hbox to 0pt{#1\hss}}
\def\ltwig{\mathrel{\spose{\lower 3pt\hbox{$\mathchar"218$}}
     \raise 2.0pt\hbox{$\mathchar"13C$}}}
\def\gtwig{\mathrel{\spose{\lower 3pt\hbox{$\mathchar"218$}}
     \raise 2.0pt\hbox{$\mathchar"13E$}}}
\def\blankline{\par\vskip \baselineskip}
\def\beq{\begin{equation}}
\def\eeq{\end{equation}}

\def\=={\equiv}

\def\dmu{\left | {d\mu \over dx}\right |}
\def\emit{{j}}
\def\emito{{j_{o}}}
\def\rp{{r_{+}}}
\def\rmin{{r_{-}}}
\def\rbar{{\bar r}}

\def\taus{{\tau_{S}}}
\def\bstar{{\xi_o}}

\documentclass[preprint]{aastex}

\begin{document}

\title{The Zeeman Effect in the Sobolev Approximation II: Split
Monopole Fields and the ``Heartbeat'' Stokes V Profile}

\author{K. G. Gayley}
\affil{Department of Physics and Astronomy, University of Iowa, Iowa City,
IA 52245}

\author{R. Ignace}
\affil{Department of Physics and Astronomy, East Tennessee State
University, Johnson City, TN 37663}

%% basic plan: first a paper just on emission lines, to see how the
%%  "big star" effect can make strong emission lines, without having
%% the source function fall too precipitously (it doesn't in Hillier
%%  simulations.  We can contrast thin lines, thick yet effectively thin,
%%  and effectively thick, all for homologous expansion, for various
%%  assumptions about the line formation mechanism.
%%  Then, this paper will be about circular polarization from those
%%  same line formation assumptions.  There's no point in the constant
%%  v case, as it has V=0, and there's no point in the axial B case,
%%  since it is inconsistent with radial winds.
%%  We'll write the V(x) paper first, and just use heuristic line formation.
%%  Later, we can go back and put in realistic line excitation.

\begin{abstract}

We calculate the circularly polarized Stokes $V(\lambda)$ profile for
emission lines, formed in hot-star winds threaded with a weak radial
magnetic field.
For simplicity, the field is treated as a split monopole
under the assumptions that it has been
radially combed by the wind, and rotation is not playing a central role.
Invoking the weak-field approximation, 
%whereby the polarization is governed by 
%the longitudinal Zeeman effect 
%using the line-of-sight component
%of the magnetic field, 
we find that the $V(\lambda)$ profile
has a characteristic ``heartbeat'' shape exhibiting multiple
sign inversions, which might
be mistaken for noise in the absence of theoretical guidance.
We also conclude that there is a tendency
for the $V(\lambda)$ profile to integrate to zero on each side of
the line separately.
%We suggest that extremely strong lines may not be ideal
%for optimizing the detectability of $V(\lambda)/I(\lambda)$, owing to
%excessive dilution by weakly polarized emission.
%Ideal lines for constraining
%the circumstellar radial magnetic fields will be those that combine
%high Lande g factors with an overall emission-line strength that is
%roughly double the background continuum.  
The overall scale of
$V(\lambda)/I(\lambda)$ is set by the ratio of the 
field strength to
the flow speed, $B/v$, characteristic of the line-forming region,
and is of the order of 0.1\%
for a wind magnetic field $B \cong {\mbox 100~G}$ at depths where the windspeed
is $v \cong 100 $ km s$^{-1}$.

\end{abstract}

\section{Introduction}

Recent advances in spectropolarimetry have ushered in a new era of
detecting and characterizing the surface magnetic fields of hot stars
(e.g., Donati et al.\ 1997; Donati et al.\ 2006; Bouret et al.\ 2008;
Hubrig et al.\ 2008; Wade et al.\ 2009).  What still lies ahead is
extending these results into the winds, where the magnetic fields
can have an important dynamical influence.  Indeed, magnetic fields of
optically thick winds, like those of Wolf-Rayet stars, cannot be seen in
the static layers owing to shrouding by the supersonic wind.

In some instances, new detections come when pre-existing observations
are subject to closer theoretical scrutiny.
In others,
theory can {\it lead} and {\it motivate} observations, 
guiding the detections
of signals that might otherwise be too weak to distinguish confidently
from noise.  It is this latter goal of theoretical support that motivates
the approach of this paper, and as a
result we adopt a rough and exploratory
treatment of the line formation and magnetic environment.

As the detection of weak ($\ltwig 100$ G) fields is the goal here, it seems
likely that such fields will be combed out into a nearly
radial configuration for strong-wind stars, perhaps transitioning into a spiral pattern at
larger radii where the flow time approaches the rotation period (Friend \&
MacGregor 1984; Ignace, Cassinelli, \& Bjorkman 1998).  
Although pockets
of locally elevated magnetic fields may lead to a different class of
field detections in the wind,
here we explore the more generic case
of a smooth, global, and largely radial field, using a
split monopole
field to grossly represent this general configuration.
Hence our goal is to determine the characteristic
circular polarization signature of a recombination line formed
in a strong wind with a split monopole magnetic field, and to use that
prediction to assist observers in making detections if such
generic fields exist, given
that the first detections will likely be a challenge to distinguish
from noise.  
We leave the issue of highly asymmetric structures to future work.

The reason that direct diagnostics of radial fields would be of value is that
they have dynamical importance whenever the local Alfven speed approaches the
local wind speed.
Thus they are important when
\beq
1 \ \ltwig \ {B \over \sqrt{4 \pi \rho} \, v} \ \cong  \ 
{B \over \mbox{111~G}} \, {r \over 10 R_\odot} \,
\left ( {v \over \mbox{100~km s}^{-1/2}} \right ) \left (
{{\dot M} \over 10^{-5} M_\odot {\mbox yr}^{-1}} \right )^{-1/2} \ ,
\eeq
where 
$B$ is the magnetic field 
and ${\dot M}$ is the mass-loss rate.
Apparently,
for the windspeeds of interest in supersonic winds, 
$\gtwig 100$ km s$^{-1}$,
there is dynamical significance when $B \gtwig$ 100~G.
Open fields with $B$ substantially in excess of  100~G would likely leave a
conspicuous imprint on the stellar spindown time (MacGregor, Friend, \&
Gilliland 1992), but the impact of somewhat weaker fields would be
more difficult to interpret,
and a more direct diagnostic of their presence is sought.

\subsection{The weak-field approximation}

Throughout we adopt the weak-field approximation (Landi Degl'Innocenti \&
Landi Degl'Innocenti 1972;  
Jefferies, Lites, \& Skumanich 1989),
whereby the 
longitudinal Zeeman shift and the circular polarization act as though they
were responding to a magnetic field that was purely the line-of-sight
component of the actual field.
Taking our wavelength scale $x$ (measured from line center) to be in units
of the Doppler shift of a fiducial velocity $v_1$, designed to be flexible
to the context of interest,
then the contribution to the Stokes $V$ parameter for some infinitesmal
region with a given magnetic field ${\vec B}$ and without regard for any
velocity gradient (that is, not invoking the Sobolev approximation) is
\beq
\label{vstatic}
dV(x) \ = \ {dI(x+\Delta x) \ - \ dI(x-\Delta x) \over 2} \ ,
\eeq
where $dI(x)/2$ is the contribution
to the specific intensity in either polarization
in the absence of any $B$ field.
Here 
\beq
\label{deltax}
\Delta x \ = \ {\Delta \lambda_B \over \lambda_o} {c \over v_1} \ = \
{{\vec B} \cdot {\hat n} \over v_1} \  \bstar
\eeq
determines the wavelength shift 
for a line-of-sight magnetic field
${\vec B} \cdot {\hat n}$, so it determines the modification to the $dI(x)$ in
each polarization.
Also, $\lambda_o$ is the wavelength of the line, and
\beq
\Delta \lambda_B = 1.4 \times 10^{-3} \AA \, g_{\rm eff} {
{\vec B} \cdot {\hat n} \over {\mbox 100~G}}
\left ( {\lambda_o \over 5500\AA } \right )^2
\eeq
is the longitudinal Zeeman shift.
%In other words, in the weak-field approximation,
%all the emission is treated by the circularly polarized longitundal
%Zeeman effect, as though
%the only $B$ field present were the line-of-sight component, and of
%course the observing direction is along that field component.
The constant $\bstar$ depends only on
line parameters, specifically
the effective Lande
factor $g_{\rm eff}$ and the line wavelength $\lambda_o$, and is
given by
\beq
\bstar = 7.7 \times 10^{-4}\,g_{\rm eff} \,
\left (\frac{\lambda_o}{5500\AA} \right ) \frac{{\rm km~s}^{-1}}{{\mbox G}}  \ .
\eeq

At this stage $v_1$ is arbitrarily chosen, but
our convention
for $\Delta x$ and $x$ will be to use for $v_1$ a
fiducial windspeed at the zone of peak line formation, a point
which we term $r=1$ because we also scale the radius to this point.
This flexible approach allows the
actual value of $x$ and $\Delta x$ to be
interpreted from the shape of
the line profile itself, rather than requiring independently specified
physical scales.
Note that the above 
expressions merely serve to describe the action of the weak-field
approximation and the role of the Zeeman shift; below we will add the assumption of
of a steep velocity gradient and apply the Sobolev approximation, modifying
eq. (\ref{vstatic}) to account for the interplay between the velocity $v$ gradient and
the $B$ gradient.

The weak-field approximation applies for fields whose Zeeman shift,
expressed in units of velocity, is much smaller than the characteristic
broadening velocity of the profile.  For unresolved surface fields, that
is the rotational velocity of the star, but for winds, it is the 
windspeed $v_1$ characteristic of the line formation.
The latter
can be much faster, and indeed our neglect of rotation requires that it is.  
Such windspeeds are extremely large on the scale of the Zeeman effect, since
a 100~G magnetic field
produces Zeeman shifts of only about 1~km~s$^{-1}$ in optical lines,
and this justifies the use of the weak-field approximation.
As such, our focus is not on strong-field stars
like Ap/Bp stars (e.g., Babel \& Montmerle 1997; Townsend \& Owocki
2005), wherein the presence of multi-kG magnetic fields leaves a rigid imprint
on the circumstellar dynamics.  Instead, we are interested in magnetic
fields that are closer to equipartition with the wind kinetic energy,
where the field dynamical importance is more subtle, but for detectable
fields, would still be important
to understand.

This approximation allows for an especially powerful conceptual convenience
in the context of {\it radial} fields, because then the line-of-sight component of
the $B$ field depends on angle in the same way as does 
the line-of-sight component of the windspeed
$v$ (also assumed radial).
As such, the Zeeman shift has an effect that
closely mimics the Doppler shift, but with different sign in
the two polarizations.  Then the presence of a radial $B$
field serves as a kind of polarization-dependent
modification to the effective wind velocity law, for purposes of line diagnostics.
Subtracting the
intensity profiles from these effectively different wind velocity laws
then gives the Stokes flux profile $V(x)$.

\subsection{Previous results}

This project was begun in Ignace \& Gayley (2003), where some conclusions
were reached about the expected overall scale of the polarization, and
the prospects for detection in the winds
of $\sim$ 100~G fields seemed challenging, but 
potentially possible with
current technology.  However, that analysis included only what we term
here the ``gradient effect'', whereby the Zeeman shifts imply that the
contributing volume at a given wavelength
penetrates to slightly lower $r$ in one
circular polarization than in the other, an effect related to why the $I(x)$ profile
slopes downward as $|x|$ increases.
%The gradient in the line emissivity then
%provides an excess circular polarization in whichever mode comes
%from the brighter depth for the wavelength in question. 
However,
the current paper will elucidate an additional equally important effect,
the ``binning effect'', which accounts for the fact that a given wavelength
bin, integrated over emission in all directions, will receive contribution
over a {\it smaller solid angle} for whichever
polarization has a Zeeman shift that {\it augments} the Dopper shift.
Or looked at in an equivalent way,
the Zeeman shifts for that polarization
would cause emission into the same solid angle to be
stretched over a wider wavelength bin, by augmenting the
Doppler effect.

When the binning effect is appropriately included, we show here that optically
thin recombination lines should be expected to have their Stokes $V(x)$
signal cancel out when integrated over wavelength
on either side of the line, red {\it or} blue.
This does not imply that the entire $V(x)$ profile cancels, however,
because the positive and negative components are 
somewhat separated in frequency, and the
good spectral resolution that is
standard in modern spectropolarimetry can resolve the positive
from the negative contributions if sufficient care is taken.
Nevertheless, the presence of 
multiple inversions in the sign of the polarization suggests
that knowing what spectral shape in $V(x)$
to expect may be helpful in distinguishing
a real signal from noise.

Also, for optically {\it thick} lines, we will explore here two new effects that
also modify the Stokes $V(x)$ profile, which both have to do with
the fact that magnetized optically thick lines 
in the Sobolev approximation have different escape probabilities,
from the local Sobolev zone,
in different directions for the different polarizations.
The new effect we term the ``angle effect'' stems from the fact
that for a given observed wavelength, the two different circular polarizations
must escape their Sobolev zones along different angles to appear at
the same observed wavelength, and
that can favor the escape of one polarizaton over the other.
The other new effect we term the ``shape effect'', whereby the gradient 
in the Zeeman shift alters the Doppler gradient that sets the shape of
the Sobolev zone, augmenting escape in some directions and reducing
it in others.
This latter effect assists
the escape of photons when the field gradient augments the velocity gradient,
and hinders escape when they offset,
and that can also favor one polarization over the other.

As we shall see,
accounting for these three new effects, one in thin lines and two more in thick,
does not generally alter the overall {\it scale} of
the circular polarization, but does alter its detailed
spectral {\it signature}.
It is hoped that knowing in advance what signature to look for
will help observers disentangle what is noise from what is signal, 
at least in the generic
context of a split-monopole field treatment.

\section{Stokes V with radial B fields for optically thin lines}

Let us first consider the case of optically thin lines (so photons created
in the line escape the Sobolev zone without scattering).
Here only one additional effect needs to be added to
Ignace \& Gayley (2003), the
``binning effect''.
Below we show that the binning effect produces an opposite net circular
polarization to the gradient effect, and this signal appears closer to
line center, so the two together yield a
characteristic ``heartbeat'' signature in $V(x)/I(x)$ that exhibits
one sign inversion in each profile wing and one more at line center.

\subsection{Basic definitions}

As we are ultimately interested only in the ratio $V/I$ and its profile
shape, we can scale both
the wavelength $x$ (from line center) and the radius $r$ by their
characteristic values at the region of peak line formation.
Thus $r=1$ is typically the photospheric radius (either at the stellar
surface or at the wind photosphere for thick winds, and we include no
emission from $r < 1$), and $x=1$ is typically
the wavelength that resonates at $r=1$, which might appear as an
edge of a ``flat top'' for profiles with a
broad peak, or more generally 
as a kind of half-width at half-maximum for
more rounded profiles.
It is not necessary to be more specific, because the scaling of both $x$ and
$r$ are arbitrary to within a constant factor -- it is only the shape 
and magnitude of
$V(x)/I(x)$ that we need to understand, 
and the shape of the $I(x)$ profile itself defines the operational meaning
of $x$.

%As it is only the ratio $V(x)/I(x)$,
%the ``degree of polarization'', that is of primary interest here,
%the precise meanings of $V(x)$ and $I(x)$ are important only to 
%within an arbitrary scale factor.
What is actually measured for an unresolved point source is the flux
density (per frequency bin and per detector area) $F(x)$, but this includes
an inverse-square dependence on the distance $D$ to the source, which is not
a variable of interest here.
We remove that dependence, and also simplify the appearance of factors of $\pi$
later on,
by defining
\beq
I(x) \ = \ {D^2 \over \pi} \, F(x) \ .
\eeq
Then $V(x)$ is defined by decomposing $I(x)$ into its contributions from opposite
circular polarizations, and subtracting the left-hand (--) from the right-hand
(+) components.
%Again, these scales are arbitrary, as our concern is primarily the ratio $V(x)/I(x)$.

We also need to specify the tight connection between wavelength $x$ and 
radius $r$, enforced by the Sobolev approximation in the presence of steep $v$ gradients,
which is controlled by the Doppler and Zeeman shifts along the direction
cosine $\mu$ between the observer and the gas parcel 
in question.
That connection for each of the + and -- circular polarizations, using
eq. (\ref{deltax}) for weak fields, is
\beq
\label{xofmu}
x \ = \ -v(r) ( 1 \mp b) \mu_{\pm} \ ,
\eeq
where $v(r)$ is the velocity scaled to the windspeed at $r=1$ (and
corresponding to $\mu = 1$ and $x=1$ if magnetic fields are neglected),
and
\beq
\label{bpm}
b(r) \ = \ 7.7 \times 10^{-4}  g_{\rm eff} \left ({\lambda_o \over
5500\AA} \right ) 
\left ( {B(r) \over {\mbox \rm 100~G}} \right ) 
\left ({ v(r) \over {\mbox \rm 100~km~s^{-1}}}
\right )^{-1} 
\eeq
comes from eq. (\ref{deltax}).
Note that when $B$ measured in G equals $v$ measured in km s$^{-1}$, 
the numerical value of
$\bstar$ gives $b$.
Since $\bstar$ depends only on the line in question, 
lines with larger $\bstar$ generate larger $b(r)$ and are more
favorable for detecting circular polarization.
The overall magnitude of $b$ is set by $B/v$, and this determines the
detectability of the signal.

%Here $g_{\rm eff}$ is the effective Lande g factor for the line and
%$\lambda$ is the wavelength of the line.
The sign conventions used are 
that $v$ is regarded as positive for radially outward
flow, and $B$ is regarded as positive
for radially outward field, 
so $b$ is positive for radially outward ${\vec B}$ also.
This means that a split monopole field, for a given polarization, has
$b$ augmenting the effective $v$ in the hemisphere where ${\vec B}$
is radially outward, and reducing
it in the hemisphere where ${\vec B}$ is radially inward.
The `+' polarization for $V(x)$
is right circular, and this is always listed as the upper sign
when two signs are present.
We will always assume $|b| \<< 1$ and work only to lowest nonvanishing order,
consistent with the weak-field approximation.

Another important function is the derivative $d\mu/dx_{\pm}$, which
controls the width of the solid-angular slice at each emitting point that will
contribute in the $dx$ wavelength bin in each polarization.
Thus $d\mu/dx_{pm}$ controls the binning of the emission (and results in
the binning effect), and is given by
\beq
\label{dmu}
\left | {d\mu \over dx_{\pm}} \right | \ = \ 
{1 \ \pm \ b(r) \over v(r)} \ .
\eeq
It is apparent that when $b > 0$, so for radially outward ${\vec B}$, 
right-hand circular polarization experiences Zeeman shifts that
enhance the Doppler effect and spread the emission from a given solid
angle over a wider wavelength bin.
This in turn tends to yield a negative $V(x)$, and oppositely for
radially inward ${\vec B}$.

Finally, we
will assume throughout the
paper that there exists a known line emissivity function $\emit(r)$, 
the rate of creation of radiant energy in the line 
per volume per solid angle (assuming isotropic creation)
at radius $r$, in either of the
two independent circular polarizations (recall that the weak-field
approximation to the longitudinal
Zeeman effect treats all emission as circularly polarized, and the 
local creation
rate in the two polarizations is indistinguishable).
Further, we assume that known emissivity is of
power-law form,
\beq
\label{emit}
\emit(r) \ = \ \emito r^{-p}  \ .
\eeq
Here $p$ may be treated as a variable to treat various types of emission
mechanisms, escape probabilities, ionization degrees, and velocity laws.

Since $\emit(r)$ is per volume, the radial integrand for the
emission scales as $r^{2-p}$,
and we assume $p > 3$ and carry the integration to $r = \infty$.
The neglect of any dependence of photon creation rate 
on polarization is a reflection
of the fact that the Zeeman shift relative to the line wavelength,
$\Delta \lambda_B / \lambda_o$, is smaller than
the scale of $b$ by the factor $v/c \sim 10^{-3}$.
Hence the effects of $b$, which redistribute polarized emission over the
line, appear long before the effects of any real difference in the
creation rate of photons in the two polarizations.

\subsection{The unipolar field result $V^*$}

Since the interest here is in radial $B$ fields and radial $v$ flows, a
great simplification is offered by adopting spherical symmetry.
The main stumbling block in using spherical symmetry with magnetic
fields is, of course, that a strictly spherically symmetric field must
be monopolar, 
so we may conclude that
any physically attainable field must break spherical symmetry.
However, by choosing a {\it split} monopole field for our study, we
preserve the maximal symmetry consistent with zero magnetic divergence,
since a split monopole field is constructed from a spherically symmetric
monopole field by reversing the field throughout one hemisphere.

Because of the close connection between split monopole fields and fields
that are truly spherically symmetric (and hence unphysical),
it is actually convenient to first calculate the 
Stokes $V(x)$, which we denote $V^*(x)$, that corresponds to a 
hypothetical spherically
symmetric field.
Despite being strictly unphysical, this approach will be seen to have
pedagogical value, and in the case when $V^*(x)$ corresponds to a monopole
field (i.e., $B \propto 1/r^2$), it can be mapped to the
split monopole result after the fact, just by accounting manually for the 
hemispheric field reversal.
Furthermore, in this paper we consider only the most favorable case where
the observer sees the star along one magnetic pole, and then mapping
$V^*(x)$ into $V(x)$ is as trivial as reversing its sign on one whole
side of line center.

These considerations allow us to compute $V(x)$ from the $V^*(x)$
in complete spherical symmetry, which is a helpful simplification.
Specifically, it
allows the line profile to be calculated using a single
integral over radius, because the emission in each radial bin $dr$ maps
into each $dx$ bin according only to the solid-angle fraction
established by $d\mu/dx$, where $\mu$ is the direction cosine to
{\it any hypothetical 
observer} that would be in position to detect that emission.
Normally we would only be interested in a {\it particular} observer,
but due to the complete spherical symmetry, we can count all emission
as being detected by an array of equivalent observers in all directions,
and merely divide by the solid angle of any particular detector to find
the actual observed profile flux.
Indeed, since we are ultimately interested only in the comparison
$V^*(x)/I(x)$, there is no need for the final step, as we may 
just as well count all emission as observed emission.

Carrying out the radial integral
for the wind emission, and for simplicity ignoring occultation and any
photospheric continuum effects (hence we are picturing
fairly strong emission lines),
the results may be written
\beq
\label{iprof}
I(x) \ = \ \int_{\rp(x)}^\infty dr \ r^2 \emit(r) \dmu_+ \ + \
\int_{\rmin(x)}^\infty dr \ r^2 \emit(r) \dmu_- 
\eeq
and
\beq
\label{vprof}
V^*(x) \ = \ \int_{\rp(x)}^\infty dr \ r^2 \emit(r) \dmu_+ \ - \
\int_{\rmin(x)}^\infty dr \ r^2 \emit(r) \dmu_- \ ,
\eeq
where $r_{\pm}(x)$ is the formation depth in each polarization
for emission in the outward radial direction at $x$.  That is,
$r_{\pm}(x)$ solves $x = v(r_{\pm}) [1 \mp b(r_{\pm})]$,
%$v(r)$ and $b(r)$ appearing in $d\mu/dx$ 
%are evaluated at the $r_{\pm}(x)$ that solves
%equation~(\ref{xofmu}) for $\mu=1$, 
and since we envision all emission
as coming from $r > 1$, whenever we would otherwise have $r_{\pm}(x) <
1$ we replace $r_{\pm}$ with unity.  Also, since we are working only to
lowest nonvanishing order in $b$, there is no need to include any magnetic
effects in the calculation of $I(x)$, so for that calculation $r_{\pm}
= r_{\pm}(B=0) = {\rbar(x)}$ comes
from $v(\rbar) = x$, and we take $b_{\pm} = 0$.

Finally, for pole-on observing of a purely radial field that switches
from outward in one hemisphere to inward in the other, we have
$V(x) =  V^*(x) x/|x|$.  For other observing angles, there will be some
intra-hemispheric cancellation that is straightforward to compute by
also carrying out the azimuthal integrations that our current symmetry
assumptions allow us to suppress.  For this analysis, we are interested
in establishing the maximum detectability threshold for a radial field,
so we adopt the most optimistic assumption of pole-on observing, and
this also simplifies the mapping from $V^*(x)$ to $V(x)$.

\subsection{Heuristic ``heartbeat'' signal for $B \propto v$}

To advance our understanding of the desired diagnostic signature, we again
imagine complete spherical symmetry and move in a
pedagogically useful but even more physically impossible 
direction, by considering
$B \propto v$.
Such a situation not only violates the global divergence-free
constraint, it even shows {\it local} 
divergence whenever $v$ induces density changes, as for spherically diverging
supersonic winds.
But for schematic purposes, this assumption
has the useful characteristic that here
the Zeeman shift will be in {\it constant proportion} to the Doppler shift,
except with opposite sign in the two polarizations (see eq.~[\ref{bpm}]).

Hence for optically thin lines, the two polarizations will act {\it exactly}
like two winds with $v(r)$ rescaled by $1 \mp b_o$, where $b_o$ is a
global constant.  Such a rescaling has the effect of rescaling the $x$
axis of the profile by this same global factor, 
and rescaling the intensity axis
by the inverse amount (as the total emission in the two polarizations
is conserved when the lines are thin enough to avoid 
polarization redistribution
during escape).
It is this extremely simple behavior that motivates our schematic assumption.

Hence in this case, direct examination of the $I(x)$ profile for that
line would immediately allow us to compute $V^*(x)$:
\beq
V^*(x) \ = \ {1 \over 2} \left \{ (1+b_o)I[(1+b_o)x] \ - \ 
(1-b_o)I[(1-b_o)x] \right \} \ .
\eeq
To lowest order in $b_o$, this gives
\beq
\label{voiapprox}
{V^*(x) \over I(x)} \ = \  
b_o \left [ 1 \ + \ \frac{d \, \ln \, I}{d \, \ln \, x} 
\right ] \ ,
\eeq
and for pole-on observers, changing sign in the rear hemisphere ($x < 0$)
gives
\beq
\label{heuristic}
{V(x) \over I(x)} \ = \ 
b_o \, {x \over |x|}  \left [ 
1 \ + \ \frac{d \, {\mbox ln} \, I}{d \, {\mbox ln}\, x}
\right ] \ .
\eeq

For a generally bell-shaped $I(x)$ profile, this would produce the
results as illustrated in Figure~1, for $b_o = 10^{-3}$ (other $b_o$ values
would generate a proportional scaling).  
Although a physically unimportant case,
the pedagogical value of $B \propto v$  is that it
provides an immediate way to see
the {\it qualitative} attributes of the $V(x)$ profile
for radial $B$ fields, given only knowledge of $I(x)$ and $b_o$,
and it suggests a useful form for expressing the
more quantitative results that follow for a more physically plausible
field treatment.
Specifically, we are introduced to the overall ``heartbeat'' shape, while the
general form of eq. (\ref{heuristic}) explicitly demonstrates the
``binning'' (first term) and ``gradient'' (second term) effects.
\begin{figure}[t]
\begin{center}
\plotfiddle{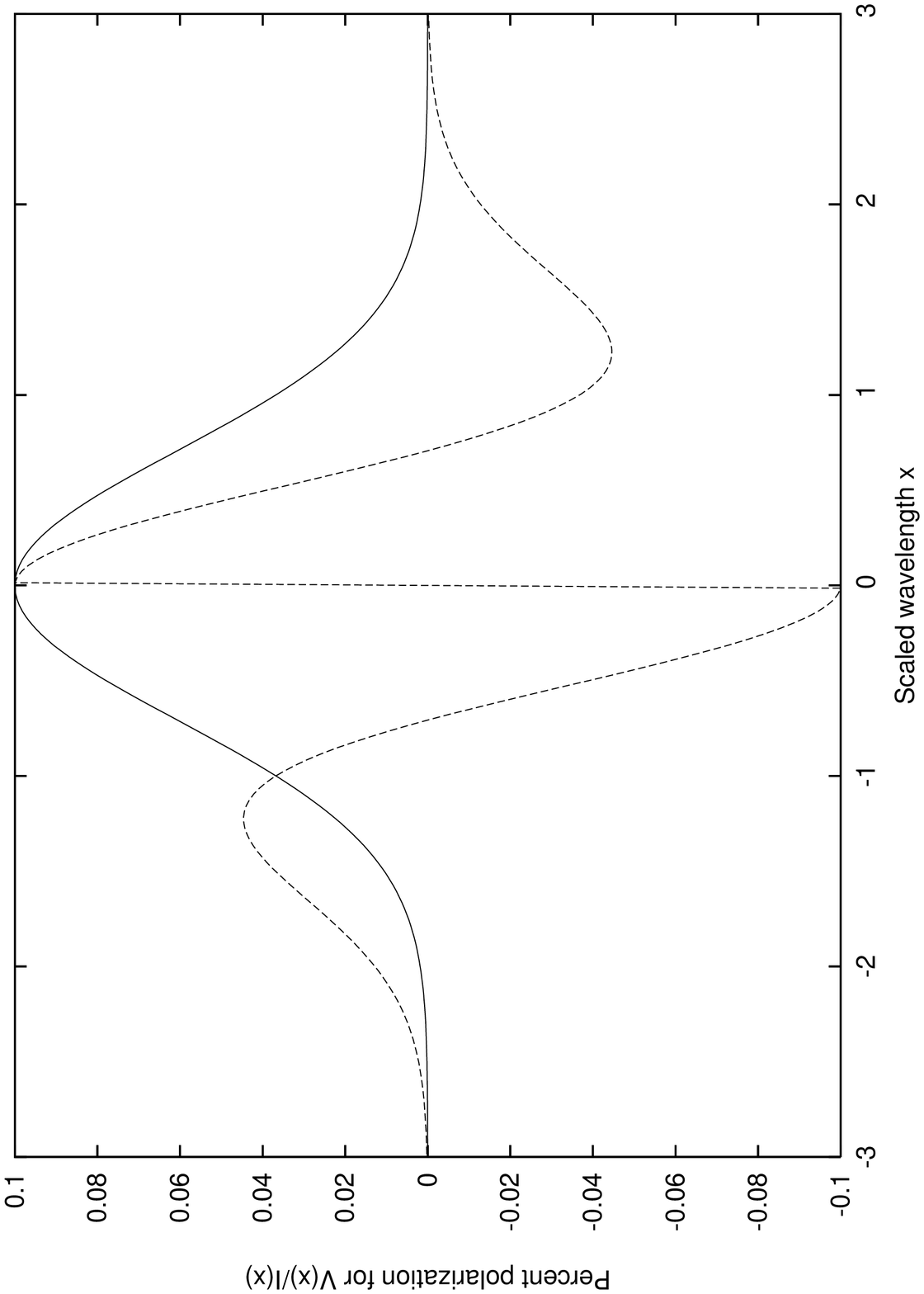}{2.6in}{-90.}{330.}{450.}{0}{0}
%\plotfiddle{PSFILE}{VSIZE}{ROTANG}{HSCALE}{VSCALE}{HTRANS}{VTRANS}
%\includegraphics[width=6.5in]{shockanglefig1.pdf}
\caption{
The percent polarization (dashed curve) of $V(x)/I(x)$ for a heuristic Gaussian profile
with a constant $b = 10^{-3}$, 
so $B$ is everywhere proportional to the radial velocity $v$, and
reverses polarity in one hemisphere, for a pole-on observer.
Although this would only be physically possible for an incompressible velocity
field, it shows the qualitative attributes of the circular polarization for an
optically thin line.
Also shown (solid curve)
is the $I(x)$ in arbitrary units, to clarify the meaning of the $x$ scale.
}
\end{center}
\end{figure}

\subsection{The split monopole results}

The simplest physically possible radial $B$ field configuration is the
split monopole, for which $B = \pm B_1 r^{-2}$ (where the ``1'' subscript
refers to the value at $r=1$), with a positive sign (outward) in one
hemisphere and a negative sign (inward) in the other.
To determine $b(r)$, we need to specify 
$v(r)$, and here we take the homologous approximation $v=r$, simulating lines
that form in the heart of the acceleration region, neither close to
nor far
from the static star.
Again the physical scale of $v$
and $r$ are arbitrary because we are interested only in the shape of
the $V(x)$ profile relative to the shape of the $I(x)$ profile.

%So if we take $v = 1$ to treat lines that form far out in
%the wind, we have $b(r) = b_1 r^{-2}$, where $b_1$ is given by eq. (\ref{bpm})
%at $r=1$, the formation radius of the line,
%and eqs. (\ref{iprof}) and (\ref{vprof}) produce
%\beq
%{V^*(x) \over I(x)} \ = \ {\int_{\rp(x)}^\infty dr \ r^{2-n} ( 1 - b_1 r^{-2} ) \ - \
%\int_{\rmin(x)}^\infty dr \ r^{2-n} (1 + b_1 r^{-2})  \over 
%2 \int_{\rbar(x)}^\infty dr \ r^{-n} } \ .
%\eeq
%Then for $x < 1 - b_1$ we have
%\beq
%r_{\pm}(x) \ =  \ \sqrt{{b_1 \over (1 \mp x) }}
%\eeq
%and $\rbar(x) = 1$, and for $x > 1 + b_1$ we have $r_{\pm}(x) = \rbar(x) = 1$.
%Carrying out the elementary integrals yields
%for $x < 1 - b_1$
%\beq
%{V^*(x) \over I(x)} \ = \
%\eeq
%with
%\beq
%I(x) \ = \   \ .
%\eeq
%Here the scale of $I(x)$ is arbitrary but its shape defines the meaning of the $x$ scale.
%%Note that simple recombination of a majority species would here yield a line emissivity
%with $n = 4$ when $v$ is constant.
%The results for $I(x)$ and $V(x)/I(x)$ for pole-on inclination with
%$n$ = 4, 5, and 6 are shown in  Fig. 2.
%%%figure 2   thin, v=1
%In the unresolved region $1 - b_1 < x < 1 + b_1$, neither expression applies and
%$V^*(x)$ varies sharply, satisfying
%an integrated constraint that we investigate below.

When we assume
homologous expansion, $v = r$, we have $b(r) = b_1 r^{-3}$,
and equations~(\ref{iprof}) and (\ref{vprof}) produce
\beq
\label{voithin}
{V^*(x) \over I(x)} \ = \ {\int_{\rp}^\infty dr \ r^{1-n} ( 1 + b_1 r^{-3} ) \ - \
\int_{\rmin}^\infty dr \ r^{1-n} (1 - b_1 r^{-3})  \over
2 \int_{\rbar}^\infty dr \ r^{1-n} } \ .
\eeq
Then for $|x| > 1$ we have for the minimum radius of formation of the
two polarizations for wavelength $x$
\beq
r_{\pm}(x) \ =  \ x \ \mp \ {b_1 \over x^2}
\eeq
and $\rbar(x) = x$, and for $|x| < 1$ we have $r_{\pm}(x) = \rbar(x) = 1$,
as we neglect line emission inside $r < 1$ to simulate a photosphere or
an ionization change.
Carrying out the elementary integrals yields
for $|x| < 1$
\beq
{V^*(x) \over I(x)} \ = \ {(p-2) \over (p+1)} b_1 
\eeq
for
\beq
I(x) \ = \  {1 \over (p-2)} \ ,
\eeq
and for $|x| > 1$
\beq
{V^*(x) \over I(x)} \ = \ -{p (p-2) \over (p+1)} b_1 x^{-3} 
\eeq
for
\beq
I(x) \ = \  {x^{-(p-2)} \over (p - 2)} \ .
\eeq
Here the scale of $I(x)$ is arbitrary but its shape defines the meaning of the $x$ scale.
It has a flat top in this case, because we assume a sharp cutoff in
the line formation at $r=1$, but only its general shape is relevant
to the discussion.
Here a simple recombination line of a majority species with $v = r$ 
would correspond to $p = 6$.
The results for $I(x)$ and $V(x)/I(x)$ for pole-on inclination
with $n$ = 4, 5, and 6 are shown in Figure~2.
%%figure 2   thin, v=r
%Again the behavior of $V^*(x)$ in the domain $1 - b_1 < x < 1 + b_1$ is constrained by an 
%integrated quantity that we discuss next.
\begin{figure}[t]
\begin{center}
\plotfiddle{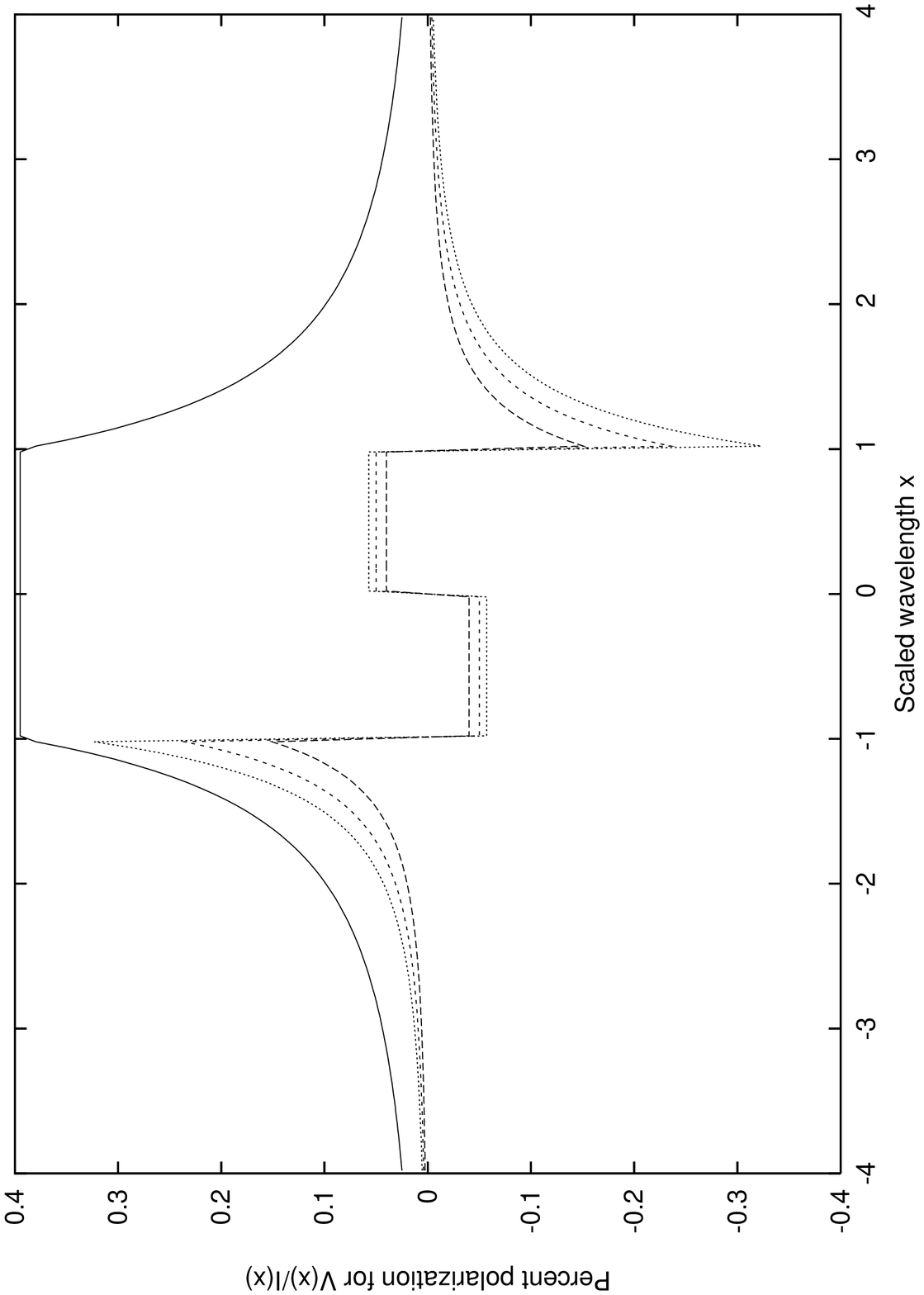}{2.6in}{-90.}{330.}{450.}{0}{0}
%\plotfiddle{PSFILE}{VSIZE}{ROTANG}{HSCALE}{VSCALE}{HTRANS}{VTRANS}
%\includegraphics[width=6.5in]{shockanglefig1.pdf}
\caption{
The percent polarization of $V(x)/I(x)$ for a thin line with power-law emissivity
where $b_1 = 10^{-3}$ at the deepest formation depth of the line, for a split
monopole field and
a pole-on observer.
The different curves are for emissivity power laws $p = 4$, 5, and 6, where the
larger $p$ yield the larger percent polarizaton, as the steeper emissivity powers
imply that more of the emission
comes from closer to the deepest formation depth $r=1$, where
$B/v$ is larger.
Also shown (solid curve)
is the flat-topped $I(x)$ in arbitrary units, to show the meaning of the $x$ scale.
}
\end{center}
\end{figure}

\subsection{A general expression for $V^*(x)/I(x)$ for thin lines}

A useful way to express equation~(\ref{voithin}) is to place it in a form similar to 
equation~(\ref{voiapprox}).
Working to lowest order in $b$ gives for $|x| > 1$
\beq
\label{vthingen}
{V^*(x) \over I(x)} \ = \  \langle b \rangle_x \ + \ 
b(x) {d \, \ln \, I \over d \,
\ln \, x}
\eeq
where $b(x)$ is $b(r)$ evaluated at $r = \rbar(x)$, and
$\langle b \rangle_x$ is an average value given by
\beq
\langle b \rangle_x \ = \ 
{\int_{\rbar}^\infty dr \ r^2 \ {\emit(r) \over v(r)} b(r) \over 
\int_{\rbar}^\infty dr \ r^2 \ {\emit(r) \over v(r)}} \ .
\eeq
Note that in the heuristic case $b = b_o$ from equation~(\ref{voiapprox}), we
have $\langle b \rangle_x = b(x) = b_o$. 
When $v = r$ and $b(r) = b_1 r^{-3}$, we have 
$\langle b \rangle_x = b_1 (n-2)/(n+1)$ 
and $d\ln I/d\ln x = 0$ for
$|x| < 1$ where the  radial integrals are from $r=1$ to $r=\infty$, and 
when $|x| > 1$, the radial integrals are from $r=x$ to $r=\infty$,
and $\langle b \rangle_x = b_1 (p-2)/(p+1)x^3$ and $d\ln I/d\ln x = 2-p$.
For split monopole fields, $b(x) = b_1/x^3$, and 
the generalized form recovers the same results as above.

Specifically, the new form shows that $V(x)$ receives contribution from a
term that depends on $I(x)$ itself, and a term that depends on the slope
of $I(x)$.  The term that is proportional to $I(x)$ stems from the ``binning''
effect, whereby for a given $x$, 
a polarization whose Zeeman shifts augment the Doppler
shift has its intensity weakened by being spread over more frequency bins,
and the term that depends on the slope of
$I(x)$ stems from the ``gradient'' effect,
whereby a polarization whose Zeeman shift augments the Doppler shift
originates from deeper depths so receives contribution from a larger volume.
The opposing signs of these two contributions suggest they
tend to cancel when {\it integrated} over either side of the emission
profile, and we next explore the simple reason why this 
integrated cancellation is
indeed complete for thin lines.

\subsection{A line-integrated constraint on V(x) for optically thin lines}

Since we consider radial fields and radial flows, and since $v = 1$
is considered to be already highly supersonic, the Zeeman shift in the
weak-field approximation does not pass any emission across line center,
so the forward hemisphere maps into the blue side of the line and the
reverse hemisphere into the red.  Then the fact that for optically
thin lines, the total emission in each polarization is the same in any
volume implies that $\int_0^\infty dx \ V(x) = 0$ when integrated over
either side of the line.  Seeing this feature in $V(x)$ profiles 
(as it continues to hold
{\it approximately}
for optically thick lines as well) would
be a good indication of the presence
of approximately radial fields, and
it underscores the need for good spectral resolution to see not only
the usual Zeeman-type overall line asymmetry, but also a polarization
reversal on {\it each side} of the line independently, which is the
general characteristic of the ``heartbeat'' feature.  

Note that it
is not $V(x)/I(x)$ that integrates to zero -- the core portion of the
percent polarization tends to be diluted by the bright core of $I(x)$, so
$V(x)/I(x)$ would tend to show more clearly the polarization in the
wings of $I(x)$.
Hence the sign of the wing polarization would dominate
if it were $V(x)/I(x)$ that was 
being integrated over $x$, as suggested in Figure~2.

\section{Stokes V with radial B fields for optically thick lines}

When the Sobolev optical depth in the line obeys
$\taus \>> 1$, photons created in the line must locally scatter 
until they are able to escape from resonance, in the familiar ways inherent in the Sobolev approximation.
This introduces two new avenues for imprinting a net 
polarization into the profile that we will
analyze next, in the context of a complete redistribution approximation.

\subsection{Complete redistribution in angle, wavelength, and polarization}

Optically thick line scattering is vastly simplified by assuming that
enough scatterings occur such that all ``memory'' of the
initial wavelength
(within the line), angle, and polarization of the newly created photon
is rapidly lost prior to
escape from the Sobolev zone.
One may then assume that each scattering is like a new creation event
for that photon, independent of its history.
This approximation has a generally
good reputation in optically thick environments, though 
its spectacularly streamlining properties are the main reason it is used.
We will make the same approximation here to take advantage of that
simplification, and
an investigation of the errors introduced is
beyond the scope of this initial investigation into the general detectability
of weak radial $B$ fields.

\subsection{The angular escape probability}

The fundamental new complication that appears for thick lines is that the 
shape of the emergent intensity profile
depends not only on the radius-dependent photon creation rate, but also
on the angle-dependent escape probability from the
Sobolev zones.
In complete frequency redistribution, the Sobolev escape probability when
$\taus \>> 1$ is inversely proportional to the Sobolev optical depth $\taus$, along the direction $\mu$
from equation~(\ref{xofmu}).
All we require is the {\it relative} escape probability $\beta(r,\mu)$
because we are treating only the {\it effectively thin} situation where all photons created in
the line escape in the line, so the relative escape probability can be normalized to unity when
integrated over escaping $\mu$.
The Zeeman shift changes this normalized relative escape probability to lowest order in $b(r)$ into
\beq
\beta_{\pm}(r,\mu) \ = \ {1 \ + \ \sigma_v \mu^2 \ \mp \ b(r) \left (1 \ + \ \sigma_b \mu^2 \right )
\over 
1 \ + \ \sigma_v/3 \ \mp \ b(r) (1 \ + \ \sigma_b/3)} \ ,
\eeq
where $\sigma_v \ = \ (d\mbox{~ln} v/d\mbox{~ln} r) - 1$ and 
$\sigma_b \ = \ (d\mbox{~ln} B/d\mbox{~ln} r) - 1$
control the line-of-sight Doppler and Zeeman shifts along $\mu$.
For $v = r$, $\sigma_v = 0$, and for split monopole fields, $\sigma_b = -3$.

As noted above, $\beta_{\pm}$ is normalized so that $\int_{-1}^{1} d\mu \ \beta_{\pm} = 1$, as 
it is a relative photon escape probability normalized over 
the (assumed isotropic) creation and scattering of photons in the line.
It already assumes an angular escape profile inversely proportional to the optical depth profile,
so applies only to lines that are optically thick to scattering, and its normalization further
requires that the line be
effectively thin in the sense
that all photons created in the line eventually escape in the line.
Effectively {\it thick} lines would not be ideal candidates for observing $V(x)/I(x)$, because they
would likely form too far out in the wind where $B/v$ is low, so would dilute the $V(x)$ signal 
against a bright and
largely unpolarized $I(x)$ profile.

\subsection{The split monopole results}

Integrating the wind emission, and ignoring occultation and any
photospheric continuum effects for simplicity,
optically thick but effectively thin lines yield the same {\it eventual} 
escape as optically thin
lines (i.e., unity), so without Zeeman influences 
they yield the same $I(x)$ as in equation~(\ref{iprof}).
However, to lowest order in $b$ we do have changes to $V(x)$ given by
\beq
\label{vprofthick}
V^*(x) \ = \ \int_{\rp(x)}^\infty dr \ r^2 \emit(r) \dmu_+ \beta_+(r,x) \ - \
\int_{\rmin(x)}^\infty dr \ r^2 \emit(r) \dmu_- \beta_-(r,x) \ ,
\eeq
where the product
\beq
\label{thickproduct}
\dmu_{\pm} \beta_{\pm} \ = \ {\left (1 \ + \ \sigma_v {x^2 \over v^2} \right )
\over \left (1 \ + \ {\sigma_v \over 3} \right ) v} \left [ 1 \ \pm \ 
{3 b \left (1 \ + \ \sigma_v \right ) 
{x^2 \over v^2} \over \left (1 \ + \ \sigma_v {x^2 \over v^2} \right )} \right ] 
\eeq
is expanded to first order in $b$, using $\sigma_b = -3$. 
Again $v$ and $b$ in these expressions are evaluated at the $r_{\pm}(x)$ that solves
equation~(\ref{xofmu}) for $\mu=1$,
and since we treat all emission as coming from $r > 1$,
whenever we would
otherwise have $r_{\pm}(x) < 1$ we
replace $r_{\pm}$ with unity.

%So if we take $v = 1$ to treat lines that form far out in
%the wind, we have $\sigma_v = -1$ 
%and $b(r) = b_1 r^{-2}$, where $b_1$ is given by eq. (\ref{bpm})
%at $r=1$, the formation radius of the line,
%and eqs. (\ref{iprof}) and (\ref{vprof}) produce
%\beq
%{V^*(x) \over I(x)} \ = \ {\int_{\rp(x)}^\infty dr \ r^{2-n} ( 1 - b_1 r^{-2} ) \ - \
%\int_{\rmin(x)}^\infty dr \ r^{2-n} (1 + b_1 r^{-2})  \over
%2 \int_{\rbar(x)}^\infty dr \ r^{-n} } \ .
%\eeq
%Then for $x < 1 - b_1$ we have
%\beq
%r_{\pm}(x) \ =  \ \sqrt{{b_1 \over (1 \mp x) }}
%\eeq
%and $\rbar(x) = 1$, and for $x > 1 + b_1$ we have $r_{\pm}(x) = \rbar(x) = 1$.
%Carrying out the elementary integrals yields
%for $x < 1 - b_1$
%\beq
%{V^*(x) \over I(x)} \ = \
%\eeq
%with
%\beq
%I(x) \ = \   \ .
%\eeq
%As before the scale of $I(x)$ is arbitrary but its shape defines the meaning of the $x$ scale.
%The results for $I(x)$ and $V(x)/I(x)$ for pole-on inclination is shown in Fig. 4.
%%%figure 4  thick line v=1 for n = 4,5,6

If we again take $v = r$ to treat lines that form in the heart of the acceleration
region where the expansion is approximately homologous, we have $\sigma_v = 0$ and $b(r) = b_1 r^{-3}$,
then equation~(\ref{thickproduct}) gives
\beq
\dmu_{\pm} \beta_{\pm} \ = \ {1 \over r} \left ( 1 \ \pm \ 3 b_1 {x^2 \over r^5} \right ) \ ,
\eeq
and equations~(\ref{iprof}) and (\ref{vprof}) now become
\beq
\label{voithick}
{V^*(x) \over I(x)} \ = \ {\int_{\rp(x)}^\infty dr \ r^{1-p} ( 1 + 3 b_1 x^2 r^{-5} ) \ - \
\int_{\rmin(x)}^\infty dr \ r^{1-p} (1 - 3 b_1 x^2 r^{-5})  \over
2 \int_{\rbar(x)}^\infty dr \ r^{1-p} } \ .
\eeq
Then for $|x| > 1$ we again have
\beq
r_{\pm}(x) \ =  \ x \ \mp \ {b_1 \over x^2}
\eeq
and $\rbar(x) = 1$, and for $|x| < 1$ we have $r_{\pm}(x) = \rbar(x) = 1$.
The integrals yield for $|x| < 1$
\beq
{V^*(x) \over I(x)} \ = \ {3 (p-2) \over (p+3)} b_1 x^2
\eeq
with
\beq
I(x) \ = \ {1 \over (p-2)}  \ .
\eeq
For $|x| > 1$, we get
\beq
{V^*(x) \over I(x)} \ = \ -{p (p-2) \over (p+3)} b_1 x^{-3}
\eeq
with
\beq
I(x) \ = \ {x^{-(p-2)} \over (p-2)}  \ .
\eeq
Note that the results for $I(x)$ are identical in the thick and thin cases, because when
$v = r$ the escape from the Sobolev zone is isotropic either way, and we are treating only
the effectively thin situation where all line photons, once created,
will escape before being rethermalized.
The results for $I(x)$ and $V(x)/I(x)$ for pole-on inclination and $p$ = 4, 5, and 6 are
shown in Figure~3.
%%figure 3   thick lines, v=r, n=4,5,6
\begin{figure}[t]
\begin{center}
\plotfiddle{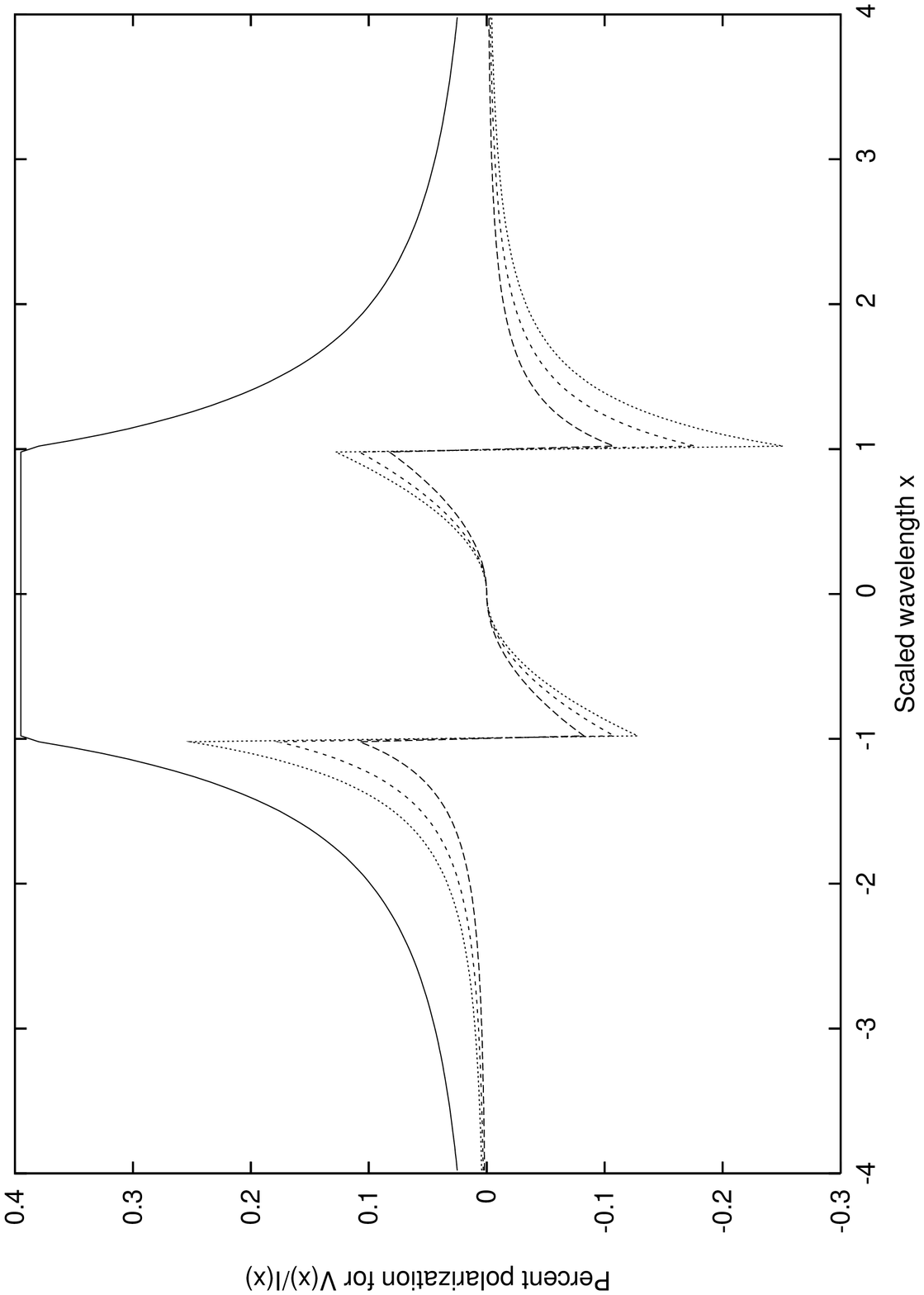}{2.6in}{-90.}{330.}{450.}{0}{0}
%\plotfiddle{PSFILE}{VSIZE}{ROTANG}{HSCALE}{VSCALE}{HTRANS}{VTRANS}
%\includegraphics[width=6.5in]{shockanglefig1.pdf}
\caption{
The percent polarization of $V(x)/I(x)$ for a thick line with power-law emissivity
where $b_1 = 10^{-3}$ at the deepest formation depth of the line, for a split
monopole field and
a pole-on observer.
The different curves are for emissivity power laws $p = 4$, 5, and 6, where the
larger $p$ yield the larger percent polarizaton just as in Figure~2.
Also shown (solid curve)
is the flat-topped $I(x)$ in arbitrary units, which again defines the meaning of the $x$ scale.
}
\end{center}
\end{figure}

\subsection{The general form for $V^*(x)/I(x)$ for thick lines with $v = r$}

The purpose of this investigation is to seek {\it general} aspects of
the circular polarization profile due to weak radial fields, and as such
the specific shape of the $I(x)$ and the $V(x)$ profiles are
not the primary concern, but rather the connection {\it between} them.
What are the attributes of the $I(x)$, given $b(x)$, that contribute
to generating a detectable $V(x)/I(x)$ polarized fraction?
To address this for thick lines,
we may again place equation~(\ref{voithick}) in a form similar to
equation~(\ref{voiapprox}), except that now
\beq
\label{vthickgen}
{V^*(x) \over I(x)} \ = \ 3 x^2  \left \langle {b \over v^2} 
\right \rangle_x \ + \ 
b(x \,) {d \, \ln \, I \over d \, \ln \, x}
\eeq
where again $b(x)$ is $b(r)$ evaluated at $r = \rbar(x)$, and
$\langle f \rangle_x$ is the same emission average as in the thin case.
If we then take $v = r$ and
$b(x) = b_1 x^{-3}$, we have for $|x| < 1$ that 
$3 x^2 \langle b/v^2 \rangle_x = 3(p-2)b_1 x^2/(p+3)$
and $d\ln I/d\ln x = 0$, recovering $V^*(x)/I(x) = -3(p-2)b_1 x^2/(p+3)$.
For $|x| > 1$, 
we have  $3 x^2 \langle b/v^2\rangle_x = 3(p-2)b_1 x^2/(p+3)$ and
$d\ln I/d\ln x = 2-p$, so $V^*(x)/I(x) = -p(p-2)b_1 x^{-3}/(p+3)$ as above.

%\subsection{Line formation models}
%   leave this to the next paper, with line models and I(x)
%Here write heuristic formation depth expressions for thin, thick, and
%effectively thick lines.

\section{Discussion}

The above results show that for thin lines,
the $V^*(x)$ profile integrates
to zero over either side of line center, but this constraint is no longer
satisfied by thick lines, because Zeeman shifts may actually favor the
net escape of the polarization that effectively augments the velocity
gradient and enhances escape in that polarization.  
However, the overall ``heartbeat'' shape appears for both thick and thin lines,
so there is still polarization reversal on each side of the line, and the
potential exists
for cancellation if the wavelength scale of the inversion 
is not resolved.
The inversion should appear somewhere near where the slope in $I(x)$ is maximal,
so observational techniques that co-add bins to increase signal must
exert caution 
in that vicinity, lest the bins so generated be too wide and
enhance cancellation.
Careful inspection of the signal near sign reversals is thus advisable, if
global radial fields are present.

One caveat in regard to the problem of unintentional
polarization cancellation should be noted.
Our schematic approach of choosing a sharp cutoff in 
the line formation at $r=1$
is related to the profile kink that separates the core and wings of $I(x)$,
and this kink is also associated with the discontinuous spike in the wing
polarization seen in Figs. 2, and 3.
This spiky behavior in the wings of $V(x)$ is not seen in Fig. 1, because
there is no sharp cutoff involved in the line formation that is implicit
in that figure.
Hence, we may conclude that the sharp and discontinuous wing spike is
an artifact of our treatment, and a smoother transition is to be expected,
unless the $I(x)$ profile 
itself transitions abruptly from flat-topped to sloping downward.
A smoother transition would also involve some cancellation near the 
discontinuity in our spiky results, so one should not interpret the peak
polarization at the spikes as being real or observationally attainable.
Nevertheless, the sign transition is the feature
of interest, and the capacity to resolve that feature 
should inform the choices made when co-adding bins.

Our results show that discontinuities in $V(x)$ can also appear
at line center, as in
the unphysical situation seen in Fig. 1
where the magnitude of $B(r)$ is proportional
to $v(r)$ (but has a polarity inversion like the split monopole), 
and also for optically thin lines with a split monopole field
as in Fig. 2.
In realistic models, the line of magnetic polarity inversion would not
appear strictly in the hemisphere perpendicular to the line of sight
to the star, so the polarization inversion would be more gradual than
we depict, but nevertheless could be very noticeable.
A gradual transition with little polarization near line center, 
on the other hand, 
happens for optically thick
lines with homologous expansion as in Fig. 3, 
because then Zeeman influences near line
center ($x=0$) experience a
compensation between the ``binning effect'' (regulated by $d\mu/dx$)
and the ``shape effect'' (regulated by $\beta$), as seen
in eq. (\ref{thickproduct}), which causes $V^*(x)$ to
tend to zero as $x$ does (like $x^2$, we find).
This in turn causes $V(x)$ to pass not only continuously through zero
at line center, but also with zero slope, in contrast to
the sharp transition in thin lines.
Note the ``gradient effect'' and the ``angle effect'' are not active
near line center, so thin lines experience only the binning effect there,
whereas thick lines experience both the binning effect and the shape effect,
affecting the rate of escape along
$\mu = 0$ into the $dx$ bin.

The more generic situation is that of Figs. 1 and 2, 
where a sharp transition
is seen-- the delicate cancellation that produces Fig. 3 is a more
specialized behavior, although of course some smoothing of the transition
will occur due to any non-ideal aspects of the magnetic geometry and
inclination.
As a result, even when $B$ is globally
smooth and radial, it is suggested that observers not discount the possibility
of polarization detections {\it near} line center,
especially when a sign reversal is seen {\it across} line 
center.

The behavior of the polarization near line center, its magnitude and
the sharpness of the sign inversion, may thus be useful
in distinguishing thick from thin lines, and homologous expansion from other
forms.
For example, the
shape of $I(x)$ itself would not suffice to distinguish
thick and thin lines under homologous expansion,
because then $I(x)$ has identical shape
whenever the lines are effectively thin, regardless of whether or not they are
optically thin, because the escape from the Sobolev zone is isotropic
in either case.
%Again, various observer binning strategies might mask
%a sharp polarization discontinuity across line center,
%if the bin at line center co-adds from both sides.

Our most robust results, which depend the least on our model assumptions, is
the general form of eqs. (\ref{vthingen}) and (\ref{vthickgen}).
These show how the shape of $I(x)$ is dependent on the shape
of $V(x)$, in the most flexible terms possible, for thin and thick lines
respectively, for split monopole fields seen pole-on.
In that situation, these expressions can 
actually be used to invert the $V(x)$ and $I(x)$
profiles into an effective $b(x)$ distribution, and thereby 
probe the $v(r)$ profile assuming a monopolar $B(r)$.

Even more generally, these points suggest that for any type of polarization
detection, considerations involving the
underlying field topology and global wind model
may dictate the effectiveness of
various data analysis strategies, and mapping that connection motivates
theoretical studies of other types of field geometries in future.
This may be particularly relevant in light of the reported detection by
Donati et~al. (2005) of circularly polarized lines in the disk of
FU~Ori, demonstrating the capability to detect the longitudinal
Zeeman effect in circumstellar environments at a sub-percent degree of
polarization.

All our results, even those for optically thick lines, assume that the lines
are {\it effectively} thin, so all photons created in the
line escape in the line.
It should be noted that
when this is not the case, the $I(x)$ will drop relative
to what our expressions give, but the $V(x)$ will drop {\it even
more}, since $V(x)$ forms at the greater 
depths where $B/v$ is largest, but that is also where
the lines would be the most effectively thick 
and rethermalization would be most severe.
This would have to be seen against an 
unfortunately bright background of $I(x)$ that
forms farther out, where the wind is less dense
and rethermalization is less of a problem, and would be largely unpolarized
due to the rapidly dropping $B/v$.
Hence it is suggested that effectively thick 
lines are to be avoided, despite yielding
very bright emission lines, because the presence of many photons is not always a
benefit when they serve only to dilute the polarized component.

Similarly, emission lines that are too weak would be diluted by unpolarized continuum
emission, which we did not include in this study to 
maintain our focus on strong lines
with maximum simplicity.
Hence it is suggested that lines of intermediate strength, with peak 
intensities about as bright as the continuum, and significant
emission from regions in the wind moving considerably slower
than 1000 km s$^{-1}$, 
would be best
suited for detecting global radial fields.
Lines with larger values of the $\bstar$ parameter are also favored.

%Although our results assume a pole-on inclination of the $B$ field,
%information may be extracted from
%the shape of the $V(x)/I(x)$ profile to infer the actual
%magnetic inclination.  For pole-on viewing of a split monopole, the $V(x)$
%is found simply by antisymmetrizing $V^*(x)$, but when the pole has an
%inclination, partial cancellation from opposite magnetic hemispheres
%will reduce the $V(x)$ near line center.  Only for large inclinations
%will the $V(x)$ be affected farther into the blue wing, because the blue
%wing comes from the most forward portions of the magnetic hemisphere.
%Magnetic field strengths that varied with polar angle could also mimic
%a similar effect, but if there were also a modulation on the rotation
%period, the effects of magnetic cancellation could be disentangled from
%the impact of gradients with polar angle.  Such considerations are beyond
%the current scope of this paper, though more detailed modeling if and when
%promising data becomes available would certainly benefit from including them.
%This would be particularly true when rotational modulation is used to
%strengthen the separation of signal from noise.

\section{Summary and Conclusions}

Figures 2 and 3, and eqs. (\ref{vthingen}) and
(\ref{vthickgen}), present the fundamental results of this paper, that the
circular polarization profile $V(x)/I(x)$ for a split monopole field
should exhibit a characteristic ``heartbeat'' shape, explained
heuristically in Figure~1.  The scale of the circular polarization
is determined by the $b$ parameter, so by the ratio $B/v$ characteristic
of the peak line formation region, and we find the peak signal in
$V(x)/I(x)$ reaches as high as $p(p-2)/(p+1)$ times $b_1$ for thin lines,
and around $p(p-2)/(p+3)$ times $b_1$ for thick lines.
Here $p$
is the power-law index in the volume emissivity, and
$p \cong 6$ for the recombination of the dominant species in a region of
nearly homologous expansion.  Thus $V(x)/I(x)$ might
be expected to peak at a level of nearly
$3 b_1$, although this may be 
somewhat optimistic because of smoothing mechanisms
that would create some cancellation in the sharpest spikes in $V(x)$.  
It was stated earlier that $b_1 \sim 8 \times 10^{-4}
g_{\rm eff}$ at 5500 \AA , when $B$ = 100 G and $v$ = 100 km s$^{-1}$ are
characteristic of the peak line-forming region designated $r = 1$,
manifested in the profile
near $x \cong 1$ where the wings of $I(x)$ begin to fall
steeply.  Hence it would not be surprising if
a local $B$ field of 100 G at a wind photosphere at 100 km s$^{-1}$
can yield a peak $V(x)/I(x)$ polarization at the 0.1 \% level under
ideal conditions.  
Inversely, the greatest challenge to the detection of
fields at the 100~G level is if the dominant contribution to the line
emission comes at radii where the wind is substantially faster than 100 
km s$^{-1}$.

All this appears to show promise that fields strong enough
to have any dynamical significance
in hot-star winds may lie
within the reach of modern instrumental
detectability, and it is our hope that these results will help assist
observers in making detections of such fields if they do in fact exist.
We find it especially significant that the Stokes $V$
signature, owing to the Doppler shifts of rapidly approaching and
receding wind hemispheres, reverses sign once on each side of the profile, rather than
doing so only at line center where Stokes $V$ profiles normally reverse in
static photospheric applications.
This gives observers a morphological constraint to look for which might
otherwise be mistaken for noise were it not expected from the theory.
Sharp inversions across line center are also possible and would be as
diagnostically useful as inversions in the wings.
The potential presence of multiple sign changes in the polarized
spectrum is a potentially
important issue to bear in mind when binning data to increase signal.

Rotational modulation of the $V(x)/I(x)$ signal would also
help separate signal from noise,
if the magnetic axis misaligns with the rotation
axis, because such a misalignment would create a periodic variation
in the inclination of the
magnetic pole.
If the fields are not seen pole-on, the conversion from $V^*(x)$ to $V(x)$
must include partial cancellation between the
side lobes toward and away 
from the direction of tilt of the magnetic pole, owing
to the reversal in the magnetic polarity across the neutral 
line of radial fields.
Hence the results given here are more fundamentally interpreted as a calculation
of the spherically symmetric template $V^*(x)$, with the
determination of the observed $V(x)$ following from geometric considerations
of the (possibly varying) inclination of the magnetic neutral line.
Here we consider only the ideal configuration to 
constrain detectability under the
most optimistic conditions--
clearly, the most pessimistic conditions would include a
viewing angle along the magnetic equator, resulting in complete
cancellation of the circular polarization at all wavelengths, 
unless the field is highly asymmetric.

We stress that the split monopolar configuration is chosen here simply to
exemplify a field with a
high degree of symmetry, and because of
its ``generic'' character in a dense radial
wind environment.
These seem like more natural assumptions for guiding 
observations, than would making
detailed fits to specific conditions intended to
match observations that do not 
currently exist.
Althouth theoretical considerations suggest that when rotation is unimportant,
approximately radial fields might thread dense winds, there 
remain ample opportunities
for real winds to deviate from quasi-spherical symmetry, especially when rotation {\it is}
dynamically important.
For example, evidence for such
symmetry breaking is present in the rapidly rotating WR star EZ CMa (St-Louis et
al. 1995; Morel et al. 1998), so it is certainly possible that magnetic
detections in hot-star winds may result from aspherical
pockets of intense emission from structures in the wind,
rather than from the smooth radial
fields considered here.
Only future observations can resolve this point.

Our current analysis has not considered stellar occultation or emission
from the stellar photosphere, as we have envisioned fairly
strong recombination lines for whom photospheric occultation and emission would
not dominate the wind emission.  
Neither would it be negligible, however, unless the lines are very strong, but as
we have seen, such lines would likely form too far out to be ideal for $V(x)/I(x)$
detections in a weak global $B$ field.
Thus, for important lines of interest,
circular polarization from the red wing may well be reduced by occultation,
and from the blue wing may be 
diluted somewhat by the photospheric continuum.
Our results are therefore not yet at the level of direct modeling of observed profiles,
such detailed modeling is better attempted once promising observational detections already exist.

Also, our simple model does not account for clumping,
despite significant evidence that clumping is prevalent in
hot-star winds (Hillier et al. 2003; Zsargo et al. 2008).
But since our approach is sensitive to the radial distribution of
$\emit$ but not its overall magnitude, only radially varying emission
enhancements due to clumping would have an impact.
Any such radial dependence could be treated as a straightforward modification
to the power-law in $\emit$, 
which is one of the reasons we parametrized the emissivity
in a fairly general way.
Since clumping that did not affect the $B$ field would 
tend to affect $I(x)$ and $V(x)$ similarly,
it would likely not have a significant impact on the degree of polarization.
However, depending on the dynamics of the clumping process, it might engender a local
increase in $B$, increasing the field detectability in pockets where the
emission is especially strong.
These are viewed as quantitative details
that must be considered at the level of profile fitting once the basic
effects are detected.  

Besides these issues, it will also be important
to expand our approaches to other geometries.  Already, we have made
an initial attempt at predicting the $V$ profile from a Keplerian disk
(Ignace \& Gayley 2008).  Further explorations of disks and
extensions to rigidly
corotating structures (e.g., relevant to Ap/Bp stars like $\sigma$ Ori E, 
Townsend, Owocki, \& Groote 2005) will be
pursued in the future.  
%Here our goal is simply to assist observers in
%making initial detections of globally smooth radial fields at the roughly
%100~G level, and we can conclude that the prospects for detecting such
%fields with current technology are reasonably good if such fields are
%present and seen approximately pole-on, in moderately
%bright wind-broadened emission lines.
%When such a signal is not seen, it would likely be due to the $B$ being significantly
%below 100 G, or the
%$v$ being significantly above 100 km s$^{-1}$, in the region of primary line formation.

We would like to acknowledge helpful conversations about the current
state-of-the-art of polarization measurements with Nicole St-Louis, Tony
Moffat, Joe Cassinelli, and Ken Nordsieck.
This work was supported by NSF Grant AST-0807664.

\blankline
\noindent{\bf References}
\blankline

\noindent 
Babel, J., Montmerle, T., 1997, A\&A, 323, 121
Henrichs, H.~F., 2008, A\&A, 490, 793

\noindent
Bouret, J.-C., Donati, J.-F., Martins, F., Escolano, C., Marcolino, W., 
Lanz, T., Howarth, I.~D., 2008, MNRAS, 389, 75

\noindent
Donati, J.-F., Howarth, I.~D., Jardine, M.~M., Petit, P., Catala, C., 
Landstreet, J.~D., et al., 2006, MNRAS, 370, 629

\noindent
Donati, J.-F., Paletou, F., Bouvier, J., \& Ferreira, J. 2005, Natur, 438, 466

\noindent
Donati, J.-F., Semel, M., Carter, B. D., Rees, D. E., Collier Cameron, A. 1997, MNRAS, 291, 658

\noindent 
Friend, D. B. \& MacGregor, K. B. 1984, 282, 591

\noindent
Hillier, D. J., Lanz, T., Heap, S. R., Hubeny, I., Smith, L. J., Evans, C. J., Lennon,
D. J., \& Bouret, J. C. 2003, ApJ, 588, 1039

%\noindent
%Hubrig, S., Scholler, M., \& Yudin, R. V. 2004, A\&A, 428, L1

\noindent 
Hubrig, S., Sch\"{o}ller, M., Schnerr, R.~S., Gonzalez, J.~F., Ignace, R.,

\noindent
Ignace, R., Cassinelli, J.~P., Bjorkman, J.~E., 1998, ApJ, 505, 910

\noindent
Ignace, R., Gayley, K.~G., 2003, MNRAS, 341, 179

\noindent
Ignace, R., Gayley, K.~G., 2008, in Clumping in Hot-Star Winds, (eds) 
W.-R.\ Hamann, A.\ Feldmeier, L.~M.\ Oskinova, 137

\noindent
Jefferies, J., Lites, B. W., \& Skumanich, A. 1989, ApJ, 343, 902

\noindent
Landi Degl'Innocenti, E. \& Landi Degl'Innocenti, M. 1972, SP, 27, 319

\noindent
MacGregor, K. B., Friend, D. B., \& Gilliland, R. L. 1992, A\&A, 256, 141

\noindent
Morel, T. S., Moffat, A. F. J., Cardona, O., Koenigsberger, G., \& Hill, G. M. 1998,
ApJ, 498, 413

\noindent
St-Louis, N., Dalton, M. J., Marchenko, S. V., Moffat, A. F. J., \& Willis, A. J. 1995, ApJ,
452, 57

\noindent
Townsend, R.~H.~D. \& Owocki, S.~P. 2005, MNRAS, 357, 251

\noindent 
Townsend, R.~H.~D., Owocki, S.~P., \& Groote, D. 2005, ApJ, 630, 81

\noindent
Wade, G.~A., Alecian, E., Bohlender, D.~A., Bouret, J.~C., Grunhut, J.~H.,
Henrichs, H., et al., 2009, in Cosmic Magnetic Fields: From Planets,
to Stars and Galaxies, IAU Sump.\ \#259, 333

\noindent
Zsargo, J., Hillier, D. J., Bouret, J.-C., Lanz, T., Leutenegger, M. A., \& Cohen, D. H. 2008,
ApJ, 685, 149

%\end{thebibliography}

\end{document}